\documentclass{JAIS}

\journal{JAIS-ID}
\vol{2023}

\received{xx xx 2023}
\published{xx xx xxxx}

\def\be{\begin{equation}}
\def\ee{\end{equation}}
\def\bea{\begin{eqnarray}}
\def\eea{\end{eqnarray}}

\usepackage{lineno}
\usepackage{amssymb}

\articletype{Technical Report}

\begin{document}

\title{\LaTeX\ Challenges for the directional dark matter direct detection}

\author{Kentaro~Miuchi,\auno{1}}
\address{$^1$Department of Physics, Kobe University, Hyogo 657-8501, Japan.}

\vspace{0.2cm}
Email address: miuchi@panda.kobe-u.ac.jp

\begin{abstract}
Directional methods have been considered to provide a solid proof for the direct detection of the dark matter.  
Gaseous time-projection-chambers (TPCs) are the most mature devices for directional dark matter searches although there still exist several challenges to overcome.
This paper reviews the history, current challenges and future prospects of the gaseous TPCs for directional dark matter searches. 
\end{abstract}

\maketitle


\newpage
\section{Directional dark matter direct detection}
One fourth of the total energy in the universe is thought to be in the form of unknown particles, the dark matter (DM)~\cite{ref:DMreview2018}. 
 The Weakly Interacting Massive Particle (WIMP) has been a leading candidate for DM since the earliest days, which has resulted in large technological and experimental efforts towards its detection. Such direct searches look for evidence of WIMP interactions with the normal matter inside the detectors and have, thus far, led to orders of magnitude improvements in cross-section limits, but no confirmed detection~\cite{ref:Marc2019}. 

The importance of exploiting directional information expected for DM arriving from the galaxy was already pointed out in the very early stages (1980s) of the direct detection studies~\cite{ref:Spergel1987}.
The motion of the solar system 
with respect to the DM halo results in an anisotropic flux of WIMPs in detectors on the Earth. Despite the technological challenges, the detection of this directional signature is considered to be the ultimate proof needed for the discovery of DM.
Here ''directional'' 
means the experimental sensitivity to the direction of the recoiling nucleus, either on an event-by-event basis, or statistically.
One would expect a clear signal of WIMP-nucleus elastic scatterings in the recoil direction distribution as a forward-backward asymmetry.

\begin{figure}[ht]
\centering
\includegraphics[width=9cm]{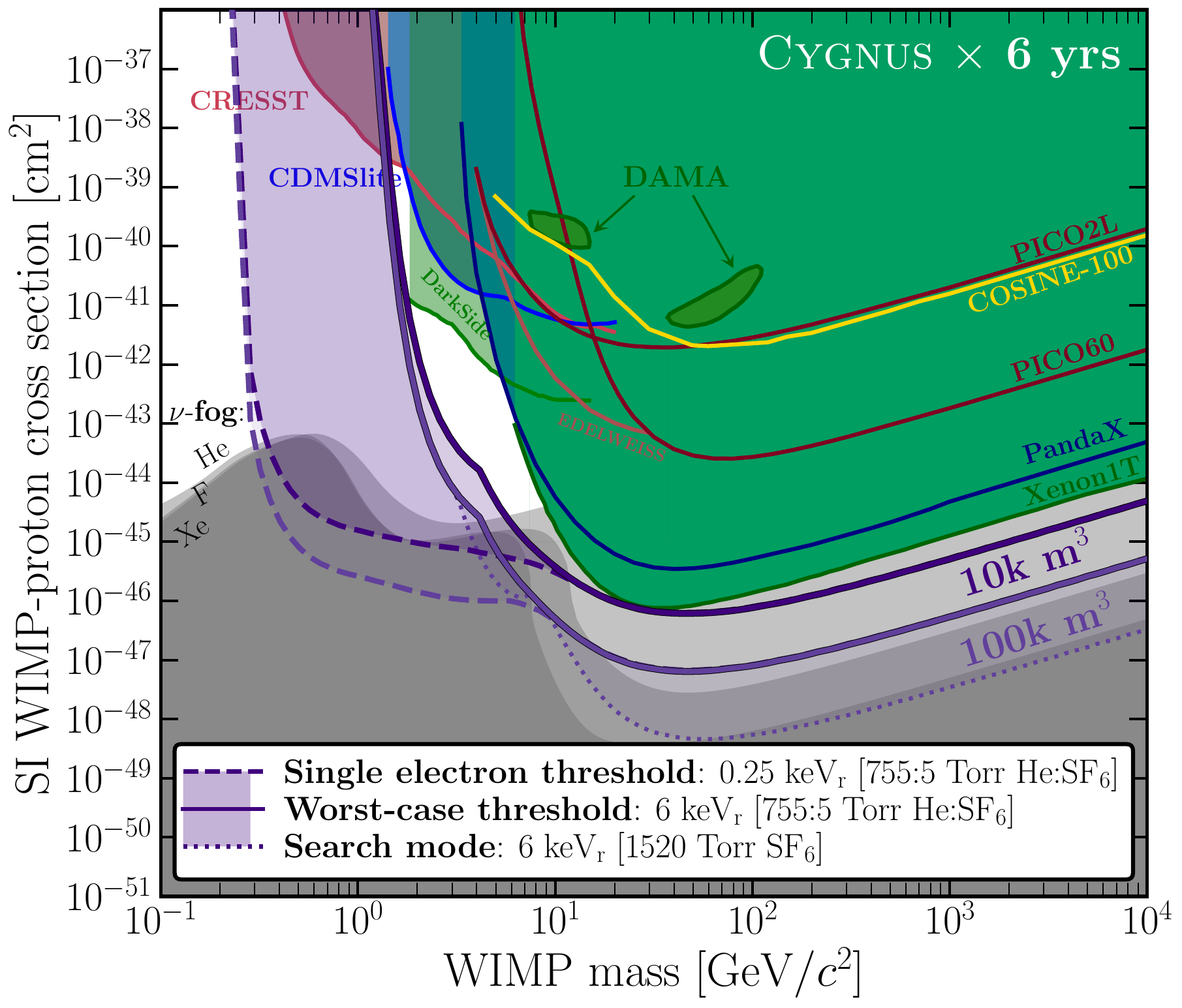}
\includegraphics[width=9cm]{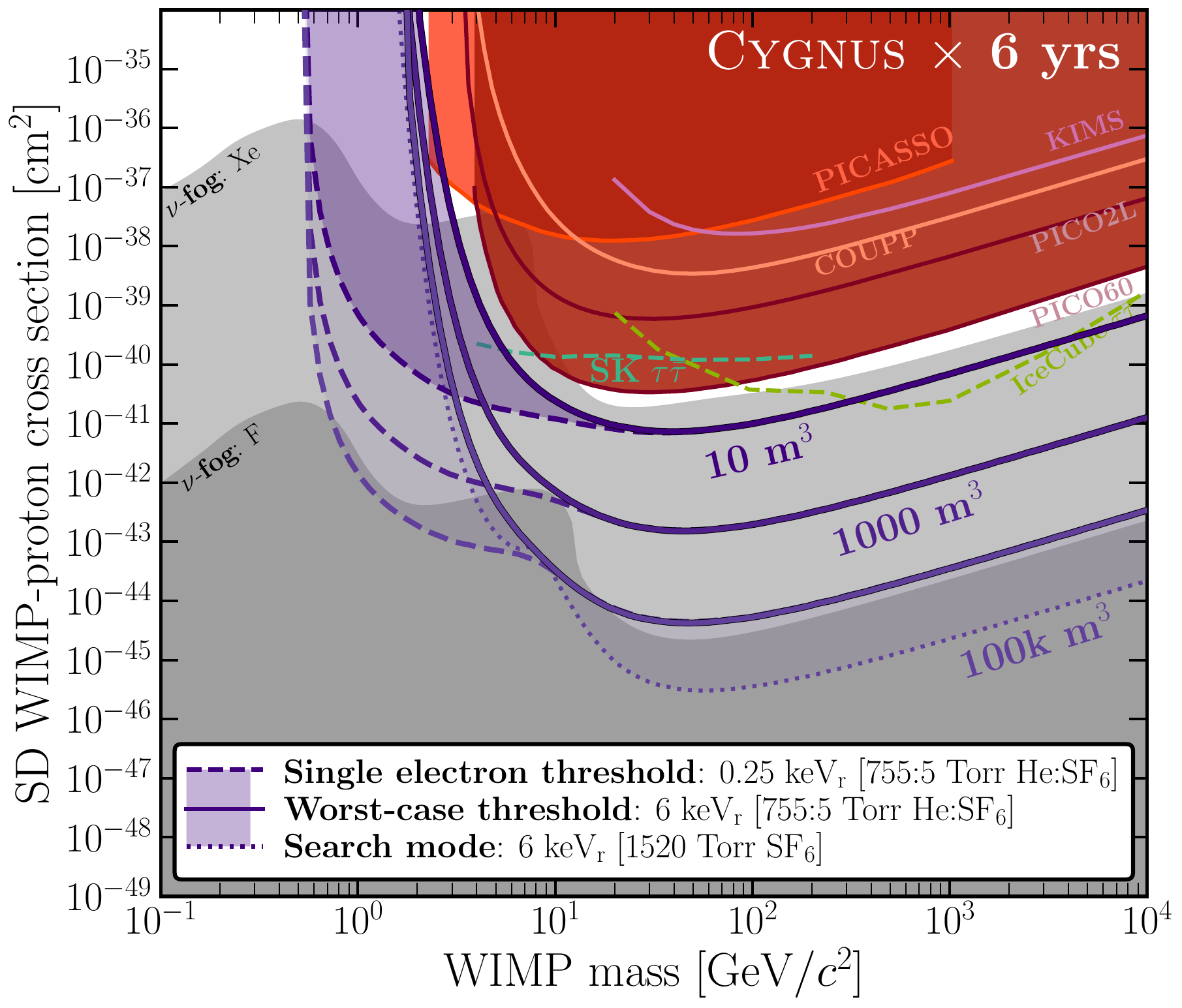}
\caption{Expected physics reach of large scale TPCs ~\cite{ref:Vahsen2020CYGNUS}. The purple lines are expected reaches by a large TPC which we call ''CYGNUS'' with a measurement time of 6 years. 
Spin-independent (SI) and dependent (SD) proton-WIMP cross section are plotted as a function of WIMP mass in the left and right panels respectively. The gray-shaded area in the lower parts of each panel is the ''neutrino-fog'' region. See text in subsection \ref{subsection:challengesfor} for a more detailed explanation of these figures.}
\label{fig:CYGNUSreach}
\end{figure}

Of the numerous directional technologies including nuclear emulsions (proposed~\cite{ref:emulsion_Natsume2007} and developed~\cite{ref:emulsion_Umemoto2019,ref:emulsion_Shiraishi2021}), anisotropic scintillators (
proposed~\cite{ref:anisorosicnti_Belli1992} and 
developed~\cite{ref:anisorosicnti_SEKIYA2003,ref:anisorosicnti_Cappella2013,
ref:anisorosicnti_Juan2020}), the columnar recombination technique (proposed~\cite{ref:columnar_Nygren_2013} and studied~\cite{ref:columnar_Nakajima_2015, ref:columnnar_Scene_2015,ref:columnar_Nakamura_2018,ref:columnar_DS20k}), and 
crystal-defect observations (proposed~\cite{ref:NVcenter2017} and studied (review)~\cite{ref:NVcenter2022}),
the gaseous time-projection-chamber (TPC) is the most well-studied and matured one.
We focus on the technological challenges of gaseous TPCs in this paper. The advantage of the TPCs for directional dark matter detection is that they can detect three-dimensional tracks of recoil nuclei, which is a clear signal of directionality. 
The largest drawback of the gaseous TPCs is their small target density ($\mathcal{O}(\rm kg/m^3$)), which is imposed by the need to operate at low pressures ($\mathcal{O}(0.1~\rm bar$)) in order to extend the track lengths of low-energy nuclear recoils above their resolution limits($\mathcal{O}(100~\mu \rm m$)). Precise numbers are given in the following part of this paper.
Detector optimization and the physics reach of the directional methods have been studied since the 2000s, {\it e.g.}~\cite{ref:Anne_2007,ref:Anne_2008,ref:Billard2011directional,ref:Vahsen2020CYGNUS,ref:OHare2021}. 
The focus of early studies was to quantify the number of events needed to detect anisotropy, which depends on the parameters characterizing tracking performance, such as the angular resolution and head-tail (track-sense) recognition. Now that the leading WIMP-searches are approaching the 
''neutrino-fog'' region 
where neutrino-nucleus coherent elastic scatterings will become the dominant, irreducible background for DM searches~\cite{ref:Billard2014neutrinofloor,ref:OHare2021,ref:Akerib2022}, interest in directional technologies has been revived~\cite{ref:OHare2015}.
The expected physics reach of large gaseous TPC experiments into the neutrino-fog are shown in FIGURE~\ref{fig:CYGNUSreach}, which we will revisit in the following sections. Directional information 
would be even more useful after the discovery of halo-WIMPs from the viewpoints of precise studies of the halo model~\cite{ref:Billard2011directional,ref:Ciaran2016,ref:Ciaran2018} and of the particle nature of WIMPs~\cite{ref:Catena_2015}.
This paper will review the technological challenges of gaseous TPCs together with their physics reach. Applications of gaseous TPCs for other physics goals and applications will also be reviewed.
Readers are also referred to previous review articles for further information, {\it e.g.}~\cite{ref:Vahsen2020CYGNUS,ref:CYGNUStechreview2010,ref:Mayet2016,ref:snowmass2021,ref:Vahsen2021DirectionalReview}. Developments and applications of TPCs for other purposes can, for example, be found in section 35 of Ref. \cite{Workman:2022ynf}.

\section{Gaseous TPCs for the directional dark matter direct detection}
\subsection{Directional signature of halo WIMPs}
The primary goal of 
the directional WIMP search is to detect the bipolar signature caused by the 
motion of the solar system with respect to the galactic halo.
Gaseous TPCs can detect and resolve the tracks of recoil nuclei on an event-by-event basis.
One would expect a clear signal of WIMP-nucleus elastic scatterings as a forward-backward asymmetry in the recoil direction distribution, 
which would be clear evidence of the halo WIMP detection.

Let us start with the fundamental physics of the 
directional signature of halo WIMPs. A spiral galaxy like our Milky Way has a rotating disk surrounded by a spherical halo. 
A ''standard'' DM halo has an isotropic Maxwell–Boltzmann distribution $f({\bf v}_{\rm gal})$,

\begin{eqnarray}
f({\bf v}_{\rm gal})&\propto&\frac{\rho_{\chi}}{m_{\chi}}e^{-\frac{1}{2}|{\bf v}_{gal}|^2/\sigma^2_0}\Theta(v_{\rm esc}-|{\bf v}_{gal}|),
    \label{eq:halovelocity}\\
{\bf v}_{\rm gal}&=&{\bf v}_{\rm lab}+({\bf v}_{\rm 0}+{\bf v}_{\odot}+{\bf v}_{\oplus}(t)),
\end{eqnarray}
where ${\bf v}_{\rm gal}$ is the WIMP velocity in the galactic frame,
$\rho_{\chi} (\rm =0.3~GeV/c^2/cm^3)$ is the local halo density, $m_{\chi}$ is the WIMP mass,
$\sigma_0$ is the velocity dispersion of the WIMPs,
$\Theta$ is the Heaviside step function,
$v_{\rm esc} (=\rm 544~km/s)$ is the escape velocity from the galaxy, 
${\bf v}_{\rm 0}=\sqrt{2}\sigma_0 =(0, 238, 0)\rm ~km/s$ is the local standard of the rest velocity at the location of the Sun,
${\bf v}_{\rm lab}$ is the WIMP velocity in the laboratory frame,
${\bf v}_{\odot}=(11.1, 12.2, 7.3)\rm ~km/s$ is the Sun’s peculiar velocity relative to the rest velocity, and
${\bf v}_{\oplus}(t)(\rm=29.8~km/s)$ is the Earth’s velocity relative to the Sun. 
The typical values recommended in Ref.~\cite{ref:Baxter2021_directsearchconvention} are shown in the vector form ($v_r$, $v_{\phi}$, $v_\theta$), where $r$ points radially inward, $\phi$ in the direction of the Milky Way’s rotation, and $\theta$ in the direction perpendicular to the galactic plane.
The characteristic directional signature of the halo WIMP is due to the rotation of the galactic disk.
The velocity of the Solar System is ${\bf v}_{\rm 0}+{\bf v}_{\odot}$, which points in the direction of the constellation Cygnus.
With another important assumption, the halo is not co-rotating with the galactic disk\footnote{This assumption is not as solid as the motion of the Solar System and some numerical simulations with baryons have indicate a co-rotating halo~\cite{ref:Ling_2010}, while another recent study indicates slow rotation~\cite{ref:Valluri_2021}. A future physics cases of these directional methods is to study the velocity distribution of the halo.}, one can expect a strong dipole feature in the incoming direction of the dark matter biased in the direction of Cygnus.

\begin{figure}[ht]
\centering
\includegraphics[height=10cm]{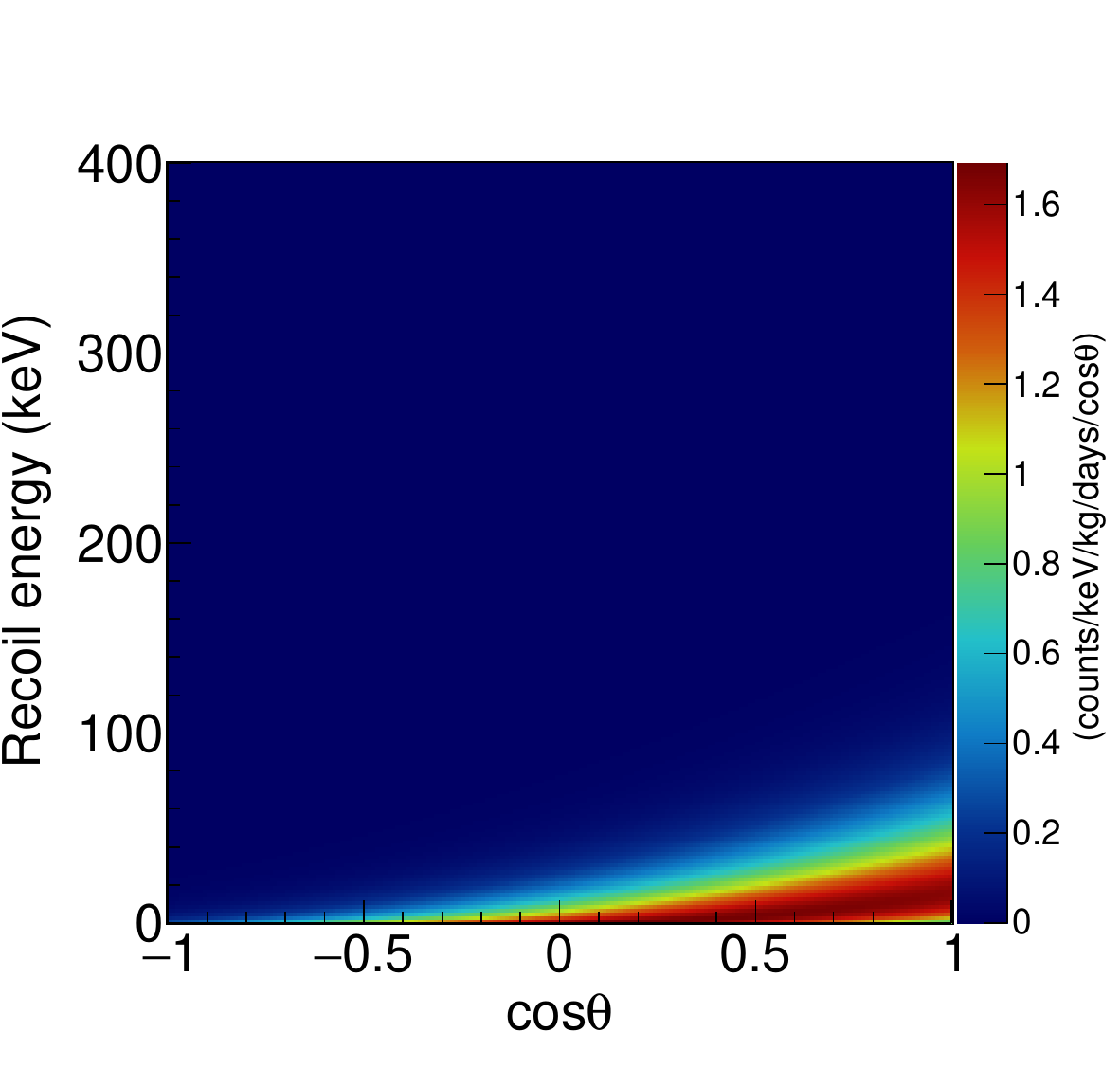}
\caption{An example of the expected energy and angular distribution of the recoil nucleus. In this plot, $\theta$ is the angle between the direction of Cygnus constellation and the direction of the recoil nuclei. Here, we assume the target to be $\mathrm{^{19}F}$ and the WIMP mass to be $m_{\chi} =100~\mathrm{GeV/c^2}$.}
\label{fig:directional2Dplot}
\end{figure}

Next, we discuss the WIMP-nucleus scattering process. Although the precise interaction is unknown, one naively expects the corresponding elastic scattering to result in forward scattering.
The direction of the recoil angle can be calculated kinematically, and a typical elastic-scattering signature is shown in FIGURE~\ref{fig:directional2Dplot}. 
The count rate is shown as a two-dimensional function of the recoil energy and the recoil angle~$\theta$ using a color map . 
Here $\theta$ is the angle between the nuclear recoil direction and the vector from Cygnus to the detector.
This angular distribution is clearly biased in the forward direction ($\cos \theta \sim 1$), an effect that becomes stronger in the higher energy region as can be understood from the kinematics. 
This intrinsic directional information is diluted by a number of 
physical properties.
The recoil nuclei undergo multiple scatterings in the detector medium.
Energy deposition along the trajectory 
decreases from the start to the end of the track, providing the physical signature needed to measure the head-tail asymmetry; it is largest at the starting point and smallest at the ending point. This trend is opposite to that expected for a typical ''Bragg curve'' because the recoil energy is low.
Ionization electrons and negative ions undergo diffusion caused by interactions with the gas molecules as they drift towards the readout device .
The original direction, total ionization and the ionization distribution along the trajectory are the measurable 
quantities used to extract the recoil direction, energy deposition and the head-tail information, respectively.

The early studies indicated that an 
isotropic distribution of the nuclear recoil can be rejected with only a few tens of WIMP events~\cite{ref:COPI1999,ref:COPI2001,ref:NEWAGE2004,ref:Morgan2005}. The next series of studies revealed that the property with the largest effect on the required number of events is the measurement of the track-sense, {\it i. e.} the head-tail of the track.
Without the track-sense information, the required number of events is increased by an order of magnitude for a readout that detects three-dimensional tracks and by two orders of
magnitude for a simplified readout that detects only two-dimensional tracks~\cite{ref:Anne_2007,ref:Anne_2008}. 
The requirement for 
constraining the parameters related to the astrophysical and particle properties of WIMPs were then discussed~\cite{ref:Billard2011directional}, followed by a thorough review article~\cite{ref:Mayet2016}. 
The discrimination of halo-WIMPs from solar boron-8 neutrinos are illustrated in Ref.~\cite{ref:Mayet2016}. 
One recent study
compared the possible technologies 
taking the cost into account~\cite{ref:Vahsen2020CYGNUS}. Details on these technologies will be reviewed in the following subsection.

\subsection{Gaseous Time-Projection-Chambers}
\label{subsection:TPC}
\begin{figure}[ht]
\centering
\includegraphics[width=12cm]{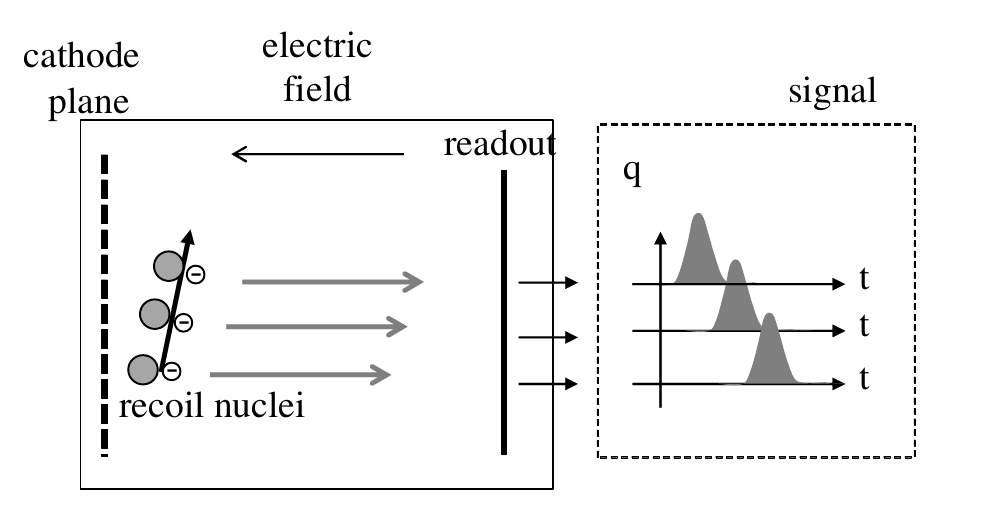}
\caption{Conceptual drawings of a TPC. Ionization electrons or ions drift to the readout, and the differences in the drift length are projected on the time differences. For a more detailed explanation of the readout schemes, please see FIGURE~\ref{fig:TPCreadouts}.}
\label{fig:TPCschematic}
\end{figure}

The inventions of the
multi-wire proportional counter (MWPC~\cite{ref:MWPC_Charpak1968}) and TPC~\cite{ref:TPC_Nygren1978} made it possible to 
detect the three-dimensional trajectories of charged particles.
A schematic drawing of a TPC is shown in FIGURE~\ref{fig:TPCschematic}. Ionization electrons or negative ions drift to the readout and the differences in their drift lengths are projected in time. Ideas for using these new technologies for the directional WIMP searches were already considered in the late 1980s~\cite{ref:lowpressureTPC1989,ref:hydrogenTPC1989} following 
theoretical calculations~\cite{ref:Spergel1987}.
We next review the study of the 
chamber gases that includes some innovative technological breakthroughs. 
A detailed collection of gas properties can be found in Ref.~\cite{ref:SAULI2022}.

\begin{figure}[ht]
\centering
\includegraphics[width=9cm]{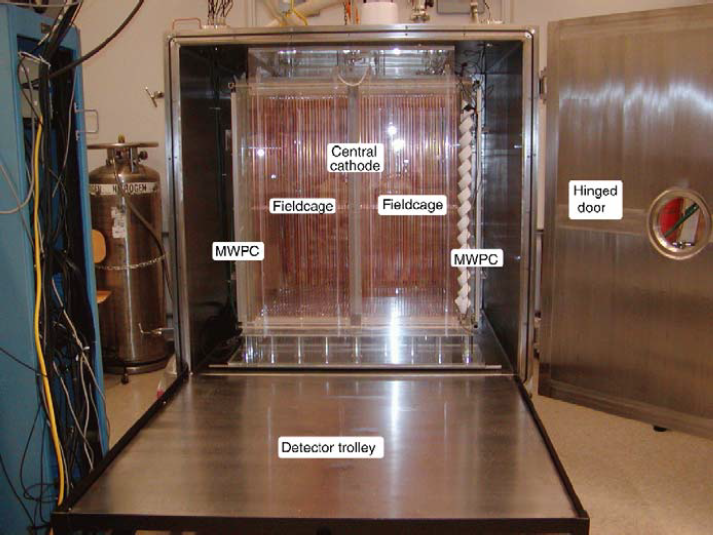}
\caption{A picture of  DRIFT-II, the first $\mathcal{O}(1~\rm m^3$)-sized detector in the field~\cite{ref:DRIFT2005_DRIFTII}. 
The central cathode is viewed by  MWPCs at both ends. 
Tubular copper rings are held by a rigid Plexiglass
support structure, forming the TPC field cage.
The detector is placed within a stainless vessel.}
\label{fig:DRIFT-II}
\end{figure}

The first impressive proof of principle was provided by nuclear track images~\cite{ref:Buckland1994}. Proton tracks in $\rm CH_4$ gas at 20~Torr and 
in a P-10 gas mixture (90$\%$ Ar and 10$\%$ $\rm CH_4$) at 50~Torr
with triethylamine (TEA) at 7$\%$ partial pressure were imaged with a charge-coupled device (CCD)~\cite{ref:Buckland1994}. 
A parallel plate avalanche chamber showed sufficiently high gas gains ($10^5$ -- $10^6$) at low pressure, and discrimination of electron tracks was also demonstrated~\cite{ref:Buckland1994}. 
Further studies focused on suppressing diffusion, which needs to be less than the typical track length over the full range of drift distances in the TPC.
Increasing the drift distance is the most cost-effective way to scale up the TPC volume, so diffusion must be kept to an absolute minimum.
The use of magnetic fields in the TPC was shown to be effective for suppressing transverse diffusion of the drifting electrons~\cite{ref:LEHNER1997}, but the added cost and complexity of large magnets rules out this option for the large detectors needed for directional DM searches. 
An important breakthrough came with the use of $\rm CS_2$, an electro-negative gas, which enabled negative-ion drift in the TPC~\cite{ref:MARTOFF2000,ref:SnowdenIfft2000}. 
In the gas mixtures of argon : $\rm CH_4$ : $\rm CS_2$ (9:1:14.5, 40~Torr) and xenon:$\rm CS_2$ (10:14.5) (40~Torr and 16.5~Torr)
ionization electrons produced by the recoiling nuclei are captured by $\rm CS_2$ molecules, resulting in the negative $\rm CS_2^{-}$ ions drifting along the TPC drift field~\cite{ref:MARTOFF2000}.
Negative ions maintain thermal drift and diffusion in all three  dimensions to much higher reduced electric fields than electrons do~\cite{ref:MARTOFF2000}. After having demonstrated the success of the negative-ion TPC, the DRIFT collaboration started  the first directional dark matter direct search experiment. The DRIFT experiment was proposed in the early 2000s~\cite{ref:SnowdenIfft2000} and a $\rm 1~m^3$-sized detector 
filled with $\rm CS_2$ at 40~Torr read by MWPC readouts (2~mm-pitched “anode wires“ and 2~mm-pitched grid-wires perpendicular to anode wires) was developed.~\cite{ref:DRIFT2005_DRIFTII}. A picture of the DRIFT-II vessel and detector is shown in FIGURE~\ref{fig:DRIFT-II}. The central cathode is viewed by  MWPCs at both ends. This bi-chamber style had often been used in many experiments mainly to double the detection volume with the same high voltage at the cathode. It was newly found by the DRIFT collaboration that the bi-chamber style with a thin cathode film helps to reduce the radioactive background and it was adopted as the standard design in the community~\cite{ref:DRIFT2015_thincathode}. One of the issues in building large-scale TPCs is the high-voltage feedthrough. 
A feedthrough made of radiopure materials for the directional dark matter searches that can withstand an operating voltage of 34~kV can be found in~\cite{ref:DRIFT2005_DRIFTII}.
It is encouraging that there exists a feedthrough that withstands an operating voltage of 100~kV
for high-energy physics ~\cite{ref:ALICE_TPC}.

The next advance came with the use of the higher resolution micro-patterned gaseous detector (MPGD) readouts and $\rm CF_4$, a "cool" (small diffusion) electron drift gas with low diffusion and a high concentration of $\rm ^{19}F$. Fluorine is a favored target for the WIMP search by spin-dependent (SD) interactions~\cite{ref:ELLIS1991}. The NEWAGE experiment was the first to explore the use of both of these technologies~\cite{ref:NEWAGE2004}.
NEWAGE performed a directional dark matter search 
with $\rm CF_4$ at 152~Torr and 
demonstrated the first use of the "sky-map" method for reconstructing nuclear recoil track directions to search for an anisotropic signature~\cite{ref:NEWAGE2007}. Here the ''sky-map'' method means an analysis method using the projected nuclear recoil track directions on the sky maps of laboratory- and galactic-coordinates. 
The $\rm CF_4$ became a ''common gas'' often mixed with other gases;  
DRIFT mixed $\rm CF_4$ with CS$_2$ (CS$_2$:$\rm CF_4$, 30:10, 40~Torr) to add SD sensitivity~\cite{ref:DRIFT_2012SD}
and the MIMAC experiment mixed CHF$_3$ with CF$_4$ (a 55~mbar mixture of 70$\%$ $\rm CF_4$ and 30$\%$ $\rm CHF_3$) in order to slow down the 
electron drift velocity of $\rm CF_4$ to match the clock rate of the electronics (50~MHz)~\cite{ref:MIMAC2013}.
Another remarkable breakthrough made by DRIFT was the 
discovery of 
multiple negative ion species with unique drift speeds, which appear when a small amount of oxygen (1 Torr) is added to the CS$_2$:CF$_4$ (30:10, 40~Torr) gas mixture~\cite{ref:DRIFTCS2}. 
This discovery enabled the determination of the z position and hence fiducialization along the drift direction. With this advance, the backgrounds originating at the cathode and at the readout plane could be rejected, which led to the first background-free DM searches in the field~\cite{ref:DRIFT2015fiducialize}. Interestingly, the $\rm CS_2-CF_4-O_2$ gas mixture was not the end of the story. This gas mixture is toxic, flammable, and explosive and a safer gas was better for underground use. 
The search for a safer alternative led to $\rm SF_6$, which was discovered to have many of the desirable properties of the DRIFT gas mixture, but without its hazards~\cite{ref:Phan2017SF6}. Negative ion drift in $\rm SF_6$ occurs at slow drift speeds and thermal diffusion, enabling $z$-fiducialization to be achieved with a small amount of $\rm SF_5^{-}$.
The drawbacks of this method are
that the minority peak made by the $\rm SF_5^-$ is smaller ($\sim 3\%$ of $\rm SF_6^-$ peak) and that the $\rm SF_6$ makes a large greenhouse effect. 
After the discovery of $\rm SF_6$, the community devoted a large effort to using this new gas, as we describe in subsection \ref{subsection:challengesby}.

\subsection{Challenges for gaseous TPCs}
\label{subsection:challengesfor}
The requirements for directional WIMP-search detectors 
have been studied for years~\cite{ref:Anne_2007,ref:Anne_2008,ref:Billard2011directional,ref:Vahsen2020CYGNUS,ref:Vahsen2021DirectionalReview}. 
Although optimization of these detectors depends on the nature of the halo WIMPs (mass, cross section, velocity distribution, ... ), 
a general consensus of the requirements has begun to form in the community~\cite{ref:Vahsen2020CYGNUS}. 
First, gaseous TPCs generally have event-level directionality with a sufficient time resolution (0.5~hours \footnote{0.5 hours of time resolution corresponds to the pointing resolution of $\sim 10^{\circ}$ for the direction of Cygnus. This resolution is sufficient when it is compared to the angular distribution shown in FIGURE~\ref{fig:directional2Dplot}. The angular distribution is broadened due to the incoming directions of the WIMPs and the elastic scatterings.}).

Two of the most important (and challenging) specifications are electron-track discrimination and head-tail (track-sense) sensitivity for nuclear tracks. Here the electron and nuclear tracks stand for the tracks of the primary electrons (mostly caused by the photo-absorptions and Compton-scatterings of gamma-rays) and nuclei (mostly caused by the scatterings of neutrons and hopefully WIMPs). We first need to select the nuclear tracks discriminating the electron tracks. We then need to know the head-tail of the nuclear tracks.
At least $\mathcal{O}(10^5)$ electron track discrimination power and a head-tail sensitivity $\geq 70 \%$ are required.
Here the electron track discrimination power is defined as the number of electron track events to 
be detected and selected as one nuclear recoil event.
The requirement for the angular resolution is modest ($30^\circ$) because the recoil nuclei retain the original direction of the incoming WIMPs through the kinematic elastic-scattering process.
These performance criteria need to be realized at a recoil  energy threshold of $\mathcal{O}(5~\rm keV_{nr})$. Here the subscript ''nr'' stands for nuclear recoil, which means the actual energy deposition by the nucleus. Similarly, the subscript ''ee'' stands for electron equivalent, which means the energy calibrated with electrons. 
The detector size should be at least 
$\rm\mathcal{O}(1~m^3)$, with the potential for scaling-up to $\rm 1000~m^3$ and even larger.

FIGURE~\ref{fig:CYGNUSreach} shows the 
expected physics reach of large volume gaseous time-projection-chambers.
The expected SI and SD WIMP-proton cross sections are shown in the left and right panels, respectively. 
The solid lines with volume labels (10~$\rm m^3$ to 100~$\rm km^3$) are directional search lines with the assumption of recoil-energy threshold of $\rm 6~keV_{nr}$. 
The neutrino-fog regions are shown in gray for helium (SI only),
fluorine, and xenon.
It is interesting to note that a large differences seen between the SI and SD neutrino-fog regions for fluorine and xenon is seen.
A He:SF$_6$ (755:5, 760~Torr) gas mixture is assumed as the chamber gas.
Helium is added to keep the chamber pressure at 760~Torr ($\sim$ atmospheric pressure) in order to make the chamber structure simple without adding any significant risk of multiple scattering. Helium also helps to enhance the SI sensitivity for low-mass WIMPs.    
The SF$_6$ gas at 5~Torr is the main target for the SD search, and a detector of $\rm\mathcal{O}(10~m^3)$ size begins to probe some part of the neutrino-fog region for xenon.

\subsection{Challenges of gaseous TPCs}
\label{subsection:challengesby}
\begin{figure}[ht]
\centering
\includegraphics[width=4.5cm]{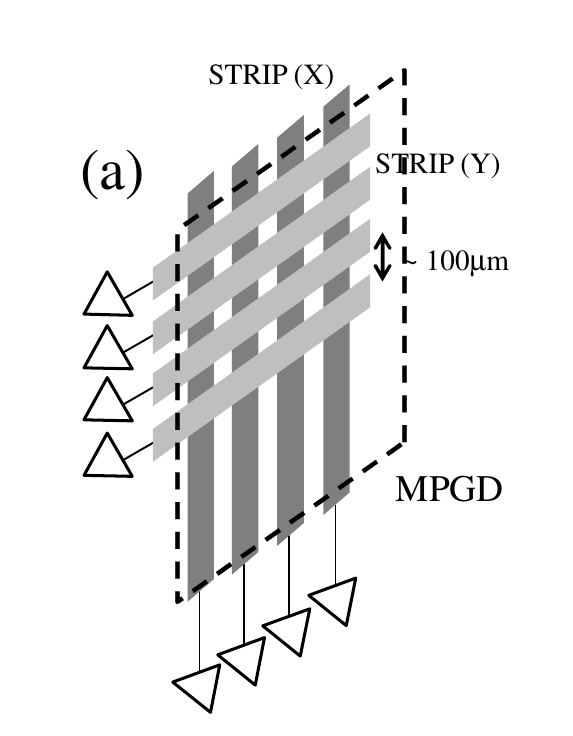}
\includegraphics[width=4cm]{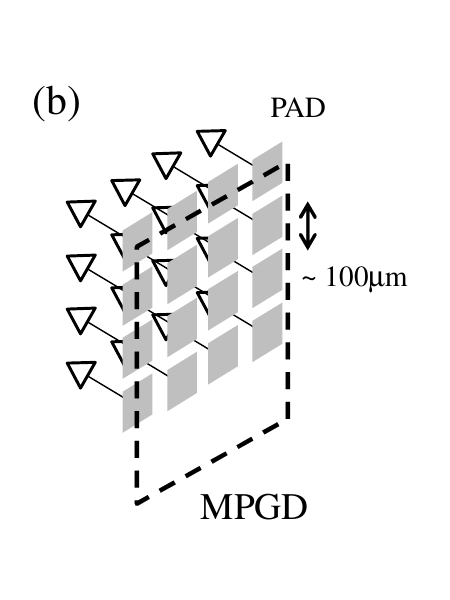}
\includegraphics[width=7cm]{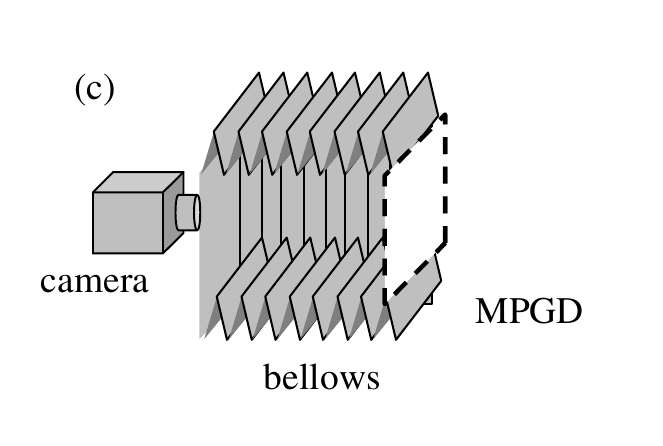}
\caption{Schematic drawings of readout systems for gaseous TPCs. Panel (a), (b), and (c) show the strip charge readout, the pixel charge readout, and the optical readout, respectively.}
\label{fig:TPCreadouts}
\end{figure}

The basic concept of a TPC vessel and detector structure for a directional WIMP search was established by the DRIFT group as shown in FIGURE~\ref{fig:DRIFT-II}.
Various readout systems are shown 
in FIGURE~\ref{fig:TPCreadouts}.
Panel (a) is a strip charge readout, (b) is a pixel charge readout, and (c) is an optical readout. The readout shown in (a) and (b) both read the charge provided by a charge amplification device, for example a MPGD. 
The difference is the number of channels, which scale $\propto L$ for (a) and $\propto L^2$ for (b), where $L$ is the size of the detector. The largest drawback of the strip  is the ''ghost'' that may appear when more than two strips on $X$ and $Y$ strips read a signal within the same time interval. In this case, both diagonal lines are equally possible directions, and they cannot be distinguished from each other. In contrast, pixel detectors have the advantages of not having ghost images at a cost of channel-number increase  proportional to the detection area ($\propto L^2$). The optical readout shown in (c) reads two dimensional optical images on the MPGDs. The largest advantage of this method is that one can use state-of-the-art commercial CCD or CMOS camera technology. The electronics is embedded within the camera itself, and the images containing up to five million pixels can be read with a single cable.
The number of cameras scales with the readout area ($\propto L^2$) with a freedom to adjust the granularity though the optical system. 
The readout is completely decoupled from the gas volume, which can be an advantage to build low  background detectors. 
Furthermore, no feed-through for data transfer or power, but only an optical window is needed.
This is particularly attractive for low pressure applications like directional dark matter detectors.
Selection of the chamber gas is limited because it needs to produce sufficient light. 
Previously, the time resolution of these cameras had not been sufficiently fast, therefore another fast readout 
device, either optical or electric, 
to obtain the timing information 
--- the third dimension --- for the tracks is required.
Some TPCs like MIGDAL/CERN use transparent ITO strips for this purpose. It should be noted that recently developed 
Timepix3-based fast camera readout demonstrated three-dimensional tracking, which will widen the possibility of using optical readout for this field ~\cite{ref:timepixRoberts_2019}.

\begin{figure}[ht]
\centering
\includegraphics[width=9cm]{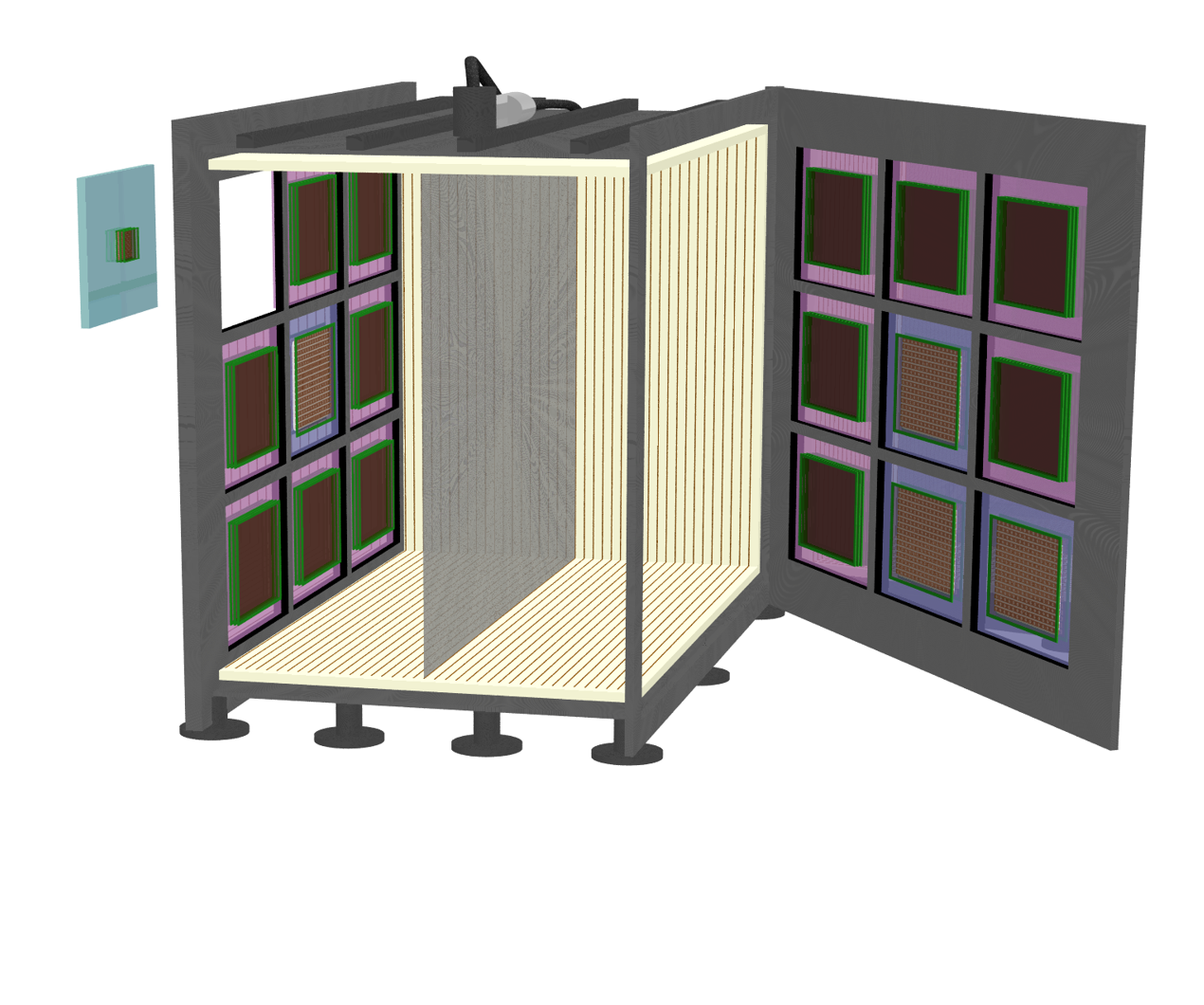}
\caption{Schematic drawing of a TPC vessel that can hold various types of readout systems. See the text for details.}
\label{fig:CYGNUS-KM}
\end{figure}

Because the nature of WIMPs is not known and the background status may vary, it is difficult to determine the ''best'' readout system. 
Combinations of different readout system may help in 
understanding of the background and identifying the WIMP signature.
The design shown in FIGURE~\ref{fig:CYGNUS-KM} is a concept for a vessel in which various types of detectors --- 18 detectors in total --- can be placed in one vessel.
While the gas handling and other vessel-related controls are shared, 
each detector has its own electronics and data acquisitions. The field cage needs to be carefully designed to ensure that the electric field is homogeneous enough even at the corners.
In the R{\&}D phase of the detectors, setting them in the same external and internal (gas-origin) background condition is beneficial to identify the detector-intrinsic background sources.
For the WIMP search, setting various detectors has advantages; some can be cheaper and can cover large part, some can be low threshold detectors and are sensitive for low mass WIMPs. 
Once some positive signals are seen, when the phase-space for the ''search'' is narrowed down, 
it is still or even more beneficial to have multiple directional detectors because  
many parameters related to the particle and astrophysical natures of the dark matter need to be known.

As we have seen in Section~\ref{subsection:TPC}, 
this field was pioneered with the MWPC technology which is categorized as a charge readout.
Using this technology, the 
DRIFT group demonstrated the operation of a large detector underground.
While DRIFT made a great improvement in the large-volume and low-background aspects, the readout technology (a 2mm-pitched MWPC) left room for improvement. The subsequent world-wide development of MPGDs produced detectors with granularities on the order of 100~$\mu$m.
Among many styles of MPGDs, gas electron multipliers (GEMs)~\cite{ref:GEM_SAULI1997}, 
Micromegas~\cite{ref:micromegas_GIOMATARIS1996}, and 
micro-pixel chambers~($\mu$-PICs~\cite{ref:uPIC_OCHI2001}) are the ones in use for this field.
The NEWAGE and the MIMAC projects pioneered the use of MPGDs for directional dark matter searches~\cite{ref:NEWAGE2004,ref:MIMAC2007}.
While MIMAC utilized Micromegas~\cite{ref:micromegas_GIOMATARIS1996}, NEWAGE used the combination of a GEM as the first stage amplifier and a $\mu$-PIC as the main amplifier and strip readout so that the amplification stage and the readout plane are partially separated and discharge countermeasure can be realized at some extent.
Both $\mu$-PIC and Micromegas have strip readouts with a pitch of $\sim 400~\mu$m~\cite{ref:TAKADA2007_uPIC30,ref:MIMAC2011}.
NEWAGE demonstrated the first directional search using a ''sky-map'' obtained from the nuclear tracks~\cite{ref:NEWAGE2007} with a $\rm 30\times30\times 30~cm^3$ detector. NEWAGE started underground measurements in 2000s and increased their search sensitivity 
while always maintaining directional sensitivity.
MIMAC started the underground measurements at the Modane underground facility in 2012~\cite{ref:MIMAC2013}. 
MIMAC developed a unique low energy ion beam facility, COMIMAC and studied directionality in detail. 
The energy range of COMIMAC is between a few tens of eV and 50~keV and the ions are transferred through a 1~$\mu$m diameter hole into the TPC~\cite{ref:MURAZ2016_COMIMAC}.
Energy calibrations in underground site can be performed with low energy X-rays from radioactive isotopes like $\rm ^{55}Fe$ and X-ray generators~\cite{ref:MIMAC2017} and alpha-particles from $\rm ^{10}B$(n, $\alpha$)$\rm ^{7}Li$ reactions~\cite{ref:NEWAGE2023}.
MIMAC resolved three-dimensional tracks of fluorine ions with energies as low as 6~keV~\cite{ref:TAO2021_MIMAC}.
They also demonstrated the head-tail recognition in the keV energy range~\cite{ref:MIMAC2022_headtail}.
The development of readout electronics is one of the important development items for gaseous TPCs because no off-the-shelf readout electronics is suited for these high density strip readouts ($\sim$25~ch/cm).
Various types application specific integrated circuits (ASIC) have been developed with bi-polar and CMOS technologies~\cite{ref:Sasaki1998_ASD,ref:APV25,ref:Orito2004_ASD,ref:RICHER2010_MIMACelec,ref:iwakichip,ref:LTARS2020,ref:LTARS2020} so far. Another important 
issue related to the electronics is the measure against discharges.
Protection diodes can be mounted for the strip readout electronics where there still some space, while more up-to-date technologies like 
diamond-like amorphous carbon~\cite{ref_DLC_ROBERTSON2002129} layer 
MPGD discharge protection~\cite{ref:Ochi_20156L} would be needed in future.

Pixel readouts, which exhibit better performance but which are technologically more challenging than strip readouts, have been developed using Timepix chips~\cite{ref:LLOPART2007_timepix} and ATLAS FE-I4 chips~\cite{ref:GARCIASCIVERES2011_FE-I4}.
Although the readout area remains limited to less than $\rm 5 \times 5~cm^2$, pixel readouts with the high-granularity of $\sim 50~\mu \rm m$ have demonstrated great potential as directional WIMP search readouts~\cite{re:Baracchini_2018_NITEC,ref:JAEGLE2019_NIM,ref:LIGTENBERG2020_gridpix}.
Precise imaging of the three-dimensional electron cloud would provide important input for validating simulation tools and theoretical models for low-energy ($\sim10$~keV) nuclear recoils~\cite{ref:JAEGLE2019_NIM}. 
The track length of the nuclear recoil has been estimated by SRIM simulations\cite{SRIM}, and recent measurements started to add 
more realistic corrections to the raw outputs of SRIM\cite{ref:MIMAC2022_headtail}.
An interesting use of such precise imaging is to determine the absolute position along the drift direction by transverse diffusion measurements, although some more studies are needed for practical uses~\cite{ref:Lewis2014}.

With the help of a low-atomic-mass helium target, a feasibility demonstration has placed a limit on the SI WIMP-proton cross-section for WIMPs lighter than 10~GeV~\cite{ref:Thorpe2018phD}.
Recent pixel readout application specific integrated circuits (ASIC) development and the possible use of ASICs originally developed for other purposes like liquid argon 
would accelerate R\&D towards their practical use in directional WIMP searches~\cite{ref:LArPIX2018,ref:HigashinoMPGD2022}.

Ever since the impressive proof of principle with an optical readout~\cite{ref:Buckland1994}, this technology has been revisited whenever technological breakthroughs have occurred in the noise rate or the frame rate.
The DMTPC group used a CCD camera for the optical readout~\cite{ref:DMTPC2008NIM}, and they presented a remarkable result to the community; in particular, they found that the head-tail recognition signal to be larger than expected~\cite{ref:DMTPC2008APP}. 
The DMTPC set directional search limits~\cite{ref:DMTPC2011}, then they explored the directionality of 140~$\rm keV_{ee}$ fluorine and 50~$\rm keV_{ee}$ helium~\cite{ref:DMTPC2017_PRD}. 
The CYGNO group started R\&D on optical readouts, mainly taking advantage of the arrival of low-noise CMOS cameras~\cite{ref:CYGNO2017,ref:Baracchini2020CYGNO,ref:CYGNO2022}. They demonstrated impressive detector performance with low energy (5.9~keV) X-rays~\cite{ref:CYGNO2019}.
They also studied the performance of a 50-liter ($\rm 33\times33\times50~cm^3$) detector ("LIME") in an above-ground laboratory, including the use of transverse diffusion to determine the absolute position along the drift direction~\cite{ref:CYGNO2023}.
The CYGNO group started underground measurements in 2022 and the technology is now mostly ready for the directional search.

Any rare-event search experiments, including  directional WIMP searches, 
require a low-background environment. In the WIMP direct search experiments, the main background sources are natural radiation such as gamma-rays, neutrons, electrons, and alpha-particles. Some of them originate from cosmic-rays, so most of these experiments are performed in underground facilities. Natural radioactive isotopes like $\rm {}^{238}U$, $\rm {}^{232}Th$, and $\rm {}^{40}K$ exist in most materials on the Earth, and they are the sources of background events originating outside of the detectors. Passive and active shields are used to shield against these external backgrounds. Directional WIMP detectors are large in volume, so one of the established and cost-effective technologies to shield them is with water shieldings. 
Simulation studies have shown that shieldings with a thickness of 75~cm of water can provide a sufficient reduction 
for a $\rm 10\times10\times10~m^3$-sized gaseous TPC~\cite{ref:Vahsen2020CYGNUS}. 
It is not possible to shield against radiation from radioactive materials in the shieldings itself, so the selection of pure materials and analytical electron discrimination are  realistic countermeasures. 
Gaseous TPCs are generally good at particle identification using linear energy transfer information obtained from the trajectory and energy deposition.  
A discrimination of $10^5$ at $\rm 5~keV$ for 
electrons was demonstrated with a strip readout~\cite{ref:MIMAC2016} and a discrimination of $>10^6$ for electrons at energies $\geq 9~\rm keV$ was demonstrated with a pixel readout~\cite{ref:Ghrear2021_JCAP}.
Discrimination of electron tracks was also measured with an optical readout; 10~$\rm keV_{ee}$ electrons were discriminated from 23~$\rm keV_{nr}$ recoil nuclei in the $\rm CF_4$ gas at 100~Torr~\cite{ref:PHAN2016APP}.
Discrimination between the X-rays from $\rm{}^{55}Fe$ and the neutrons from an Am/Be source 
with an nuclear track detection efficiency of 18$\%$ and electron track discrimination power of 96$\%$ 
was demonstrated with an optical readout~\cite{ref:Baracchini2021_CYGNO}.
Common problematic background sources in the liquid and gaseous detectors are the 
radioisotopes $\rm{}^{222}Rn$ and $\rm{}^{220}Rn$.
They are in the decay chains of $\rm{}^{238}U$ and $\rm{}^{232}Th$, respectively. The radon isotopes are rare gases that emanate from these materials.
They contaminate the gas, and the progenies of the radon isotopes 
are background sources as well.
DRIFT found that the radon progenies that accumulated on the cathode plane were background sources when they deposited part of their energies in the detection volume~\cite{ref:DRIFT_BGstudy_APP2007,ref:DRIFT2014BGstudy}. 
Using a thin film for the cathode plane and vetoing by using the counterpart detector proved to be an effective way to actively reduce these backgrounds~\cite{ref:DRIFT2015_thincathode}. Another, more general, approach is to use radio-pure materials for the detector components ~\cite{ref:hashimoto2020nim}.
Radon molecules can be captured by so-called molecular sieves which have pores designed to match the size of the radon atom. Molecular sieves made of radio-pure materials have now been developed for radon filtering~\cite{ref:Gregorio2020MS}.
$Z$-fiducialization is another way to reject the background events by radon progenies as we have seen in subsection \ref{subsection:TPC}. One caveat we should keep in mind for the future development is that not only $\rm SF_6$ (global warming potential in 20 years ($\rm GWP_{20}$)=17500)  but also $\rm CF_4$ ($\rm GWP_{20}$=4880) is a serious global warming gas~\cite{ref:GWPvalues}. Therefore we need to find replacement gases or at least circulate the gas and not use gas systems open to the atmosphere. Development of low background filtering materials is also important to keep the gas quality in a closed system during a long-term measurement.

Head-tail recognition was first demonstrated by the begging project~\cite{ref:DMTPC2008NIM}, followed by DRIFT~\cite{ref:Battat2016_headtail}.
NEWAGE used head-tail in three-dimensional tracks and applied this technique for a directional WIMP search~\cite{ref:yakabe2020ptep}.
Head-tail recognition in the keV energy range was demonstrated by MIMAC~\cite{ref:MIMAC2022_headtail}. The energy deposition of the recoil nuclei of below several tens of keV decreases along it trajectory
(larger dE/dx at the beginning and smaller dE/dx around the stopping part, see Fig.7 in Ref.~\cite{ref:DMTPC2008NIM} for instance.). If we can detect this dE/dx shape either in X-Y (as a spatial image) or Z (as a time evolution), the head-tails can be measured. A non-zero (statistically better than 50:50) head-tail measurement for 38~keV fluorine~\cite{ref:DRIFTheadtail2016} and 13~keV proton~\cite{ref:MIMAC2022_headtail} have been reported, leaving a lot of room for improvements.

Simulation studies are playing very important roles in this field like in other high energy particle physics fields. 
Each experimental group develops simulations 
for their detectors and readout electronics.
Various existing tools are used; Geant4~\cite{AGOSTINELLI2003250} as a basic structure for the detector related simualtion in most of the case, 
Degrad\cite{ref:degrad} for low energy ($\sim$keV) electrons tracks, 
SRIM~\cite{SRIM} for nuclear tracks, 
Magboltz~\cite{ref:magboltz} for electron and negative ion~\cite{ref:Ishiura_2020_NIsim} transportations,  
and Garfield++~\cite{ref:garfield} for gas avalanche.

We have so far reviewed the challenges of gaseous TPC readouts for directional WIMP searches. An important recent study concerns the 
optimization of the readout. The ''best'' detector depends on the maturity of each technology. Currently, strip readouts are the most cost-effective technology to use for the directional search, while pixel readouts always provide the highest performance~\cite{ref:Vahsen2020CYGNUS}.

\subsection{Future challenges}
\begin{figure}[ht]
\centering
\includegraphics[width=15cm]{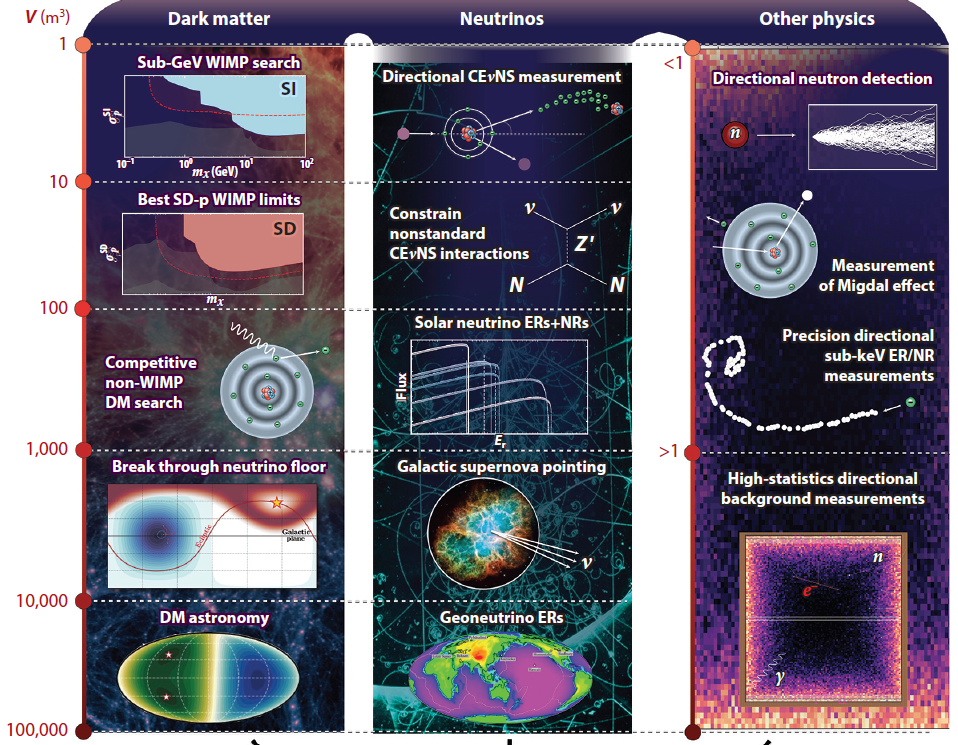}
\caption{ 
Physics cases and applications of gaseous TPCs~
\cite{ref:Vahsen2021DirectionalReview}.}
\label{fig:CYGNUSscopes}
\end{figure}

As a final topic in this review, we discuss various physics cases and applications of gaseous TPCs. 
FIGURE~\ref{fig:CYGNUSscopes} shows the physics cases horizontally and the typical required size vertically.

Some applications are already practical with detector sizes much less than $\rm \mathcal{O}(1~m^3)$.
Neutron imaging, shown in the right-top area of FIGURE~\ref{fig:CYGNUSscopes}, was realized by detecting  recoil helium tracks with pixel-readout TPCs using a detection volume of $\rm 2.0 \times 1.68 \times 10.87~cm^3$~\cite{ref:Hedges2022NIM,ref:SCHUELER2022NIM}.
Another application already in use is low-rate surface alpha-particle measurements. Surface alpha-particle counters in the market are currently being used for the industrial semiconductor impurity assay systems as well as for surface assays of materials for low-background experiments~\cite{ref:ultra-lo}.
Gaseous TPC technology can also add imaging ability to alpha-particle measurements.
A few groups are developing these devices, and they have achieved sensitivities of 
$\rm {\mathcal O}(10^{-3}~\alpha/cm^2/hour)$ with a size of $\rm (30~cm)^3$~\cite{ref:AICHAM2020,ref:AlphaCAMM2022,ref:surfacealphaMM2022}. 
The Migdal effect, which would take place by 
the quick motion of the nucleus 
as the electrons lag behind the nucleus~\cite{ref:MIGDAL_1941}, is another target for small-sized gaseous TPCs.
Among several channels of the Migdal effect, the one 
associated with nuclear recoil is attracting interest these days because it effectively lowers the energy threshold of WIMP detectors~\cite{ref:migdal_IBE_2018}. 
Several direct search experiments have extended the search region to the sub-GeV range~ \cite{XENON1T_migdal,LUX_migdal,EDELWEISS_migdal,CDEX_migdal,SENSEI_migdal,LAr_migdal}, although the effect itself is still to be observed. 
A few experimental efforts plan to observe the Migdal effect with gaseous TPCs~\cite{ref:MICALUE2020,ref:MIGDAL2023}. This can be realized with an $\rm \mathcal{O}(1~m^3)$ detector with relatively high-granularity readouts. Search for the Sub-GeV WIMPs undergoing SI interactions and SD directional searches can also be started with relatively small-sized detectors like 
$\rm \mathcal{O}(1~m^3)$--$\mathcal{O}(10~m^3)$.
It is noteworthy that a $\rm 10~m^3$ detector with low pressure SF$_6$ gas can start to search some part of the neutrino-fog region for the searches with xenon nuclei (see FIGURE~\ref{fig:CYGNUSreach}). 
Directional coherent elastic neutrino-nucleus scatterings (CE$\nu$NS) detection is also possible with detectors having sizes $\rm  \mathcal{O}(1~m^3)$ -- $\mathcal{O(10~m^3)}$.
When the technology is ready to make $\rm \mathcal{O}(100)~m^3$ -- $\mathcal{O}(1000~m^3)$ -sized chambers at a reasonable cost, these detectors would give competitive sensitivity in the electron channel.
WIMP searches beyond the neutrino-fog in the SI channel can be started with this scale. 
Solar neutrino detection can also be carried out.

$\rm \mathcal{O}(10^4~m^3)$ detectors are our ultimate goal, as they would provide a halo observatory capable of studying the 
astrophysical and particle aspects of the dark matter.

\section{Conclusions}
Gaseous time-projection-chambers are the most mature devices for directional dark matter searches, and we expect them to provide a clear detection of halo WIMPs, if they exist, and enable their precise study.
R{\&}D efforts have demonstrated component-based requirements for the directional detectors.
Detectors with a detection volume of $\rm \mathcal{O}(1m^3)$ have been operated underground for years.
MPGDs with a pitch of $\sim 400~\mu$m have also been used underground demonstrating the "sky-map" method as directional detectors.
Detection of three-dimensional tracks with non-zero head-tail sensitivity 
for low-energy ($\sim10$~keV) nuclear recoils was demonstrated. 
A discrimination $>10^6$ of low-energy ($\sim10$~keV) electrons was demonstrated.
Challenges in the upcoming years are how to integrate, including some compromise, these components to achieve intermediate goals
toward a gigantic ideal detector $\rm \mathcal{O}(10^4~m^3)$ for directional detection of WIMPs.

\section*{Acknowledgments}
KM would like to thank Dinesh~Loomba for  
useful discussions and suggestions 
throughout the paper preparation.
This work was partially supported by KAKENHI Grant-in-Aids (19H05806, 19684005, 23684014, 26104005, and 21H04471), DMNet, and ICRR joint-usage.
We would like to thank our colleagues of CYGNUS steering members ( E.~Baracchini, G.~J.~Lane, N.~J.~C.~Spooner, and S.~E.~Vahsen),   
C.~A.~J.~O'Hare, and D.~Snowden-Ifft for useful discussions and providing materials.

\bibliographystyle{bibi}
\bibliography{biblio}

\end{document}